# Mixed Fe-Mo carbide prepared by a sonochemical synthesis as highly efficient nitrate reduction electrocatalyst

*Jiajun Hu, Silvio Osella, Eduardo Arizono dos Reis, Anelisse Brunca da Silva, Caue Ribeiro, Lucia Helena Mascaro, Josep Albero and Hermenegildo Garcia\**

J. Hu, E.A. dos Reis, A.B. da Silva ,Dr. J. Albero, Prof. H. Garcia

Instituto deTecnología Química,CSIV-UPV, Av. De los Naranjos s/n, 46022 Valencia, Spain

E-mail: hgarcia@qim.upv.es

Prof. S. Osella

Chemical and Biological Systems Simulation Lab, Centre of New Technologies, University of Warsaw, Warsaw, 02–097 Poland

E.A. dos Reis, Prof. C. Ribeiro

Brazilian Agricultural Research Corporation. Parque Estação Biológica - PqEB, s/n. Postcode 70770-901 / Brasília, DF, Brazil

A.b. Da Silva, Prof. L.H. Mascaro

Departamento de Química, Universidade Federal de São Carlos, C.P. 676, 13565-905 São Carlos, SP, Brazil.

**Abstract**: Ammonia, a versatile compound that can be used as a fertilizer, chemical or fuel, has since long been produced through the energy-intensive Haber-Bosch process. Recently, the electrochemical nitrate reduction reaction (NO$_3$RR) using electricity generated from renewable sources has attracted widespread attention. However, the complex reaction pathway of NO$_3$RR leads to the formation of many undesirable by-products. Herein we successfully prepared a mixed (FeMo)$_2$C catalyst with good electrocatalytic NO$_3$RR, having a NH$_3$ yield of 14.66 mg h$^{-1}$ cm$^{-2}$ and an FE of 94.35 % at low potential -0.3 V vs RHE. DFT calculations show that the presence of Fe in Mo$_2$C lattice changes the reaction mechanism, decreasing the potential barrier to be overcome from 1.36 to 0.89 eV. In addition, mixed Fe-Mo carbide facilitates the adsorption of intermediates and promotes NH$_3$ desorption, facilitating NO$_3^-$ reduction to NH$_3$. In addition, (FeMo)$_2$C was used as cathode for Zn-NO$_3$ battery to generate electricity, producing ammonia at the same time, with a power density of 3.8 mWcm$^{-2}$ and an NH$_3$ FE of 88 %. This work describes a new synthesis method for mixed metal carbides and provides a promising strategy for NH$_3$ production.

## 1. Introduction

Ammonia (NH$_3$) is among the chemicals produced in the largest quantities. In addition to its established use as fertilizer and in the chemical industry, novel potential large scale application related to its role as carbon-free transportation fuel or liquefiable hydrogen storage chemical are being seriously considered. The energy density (4.32 kWh L$^{-1}$) comparable to that of hydrocarbons and its high hydrogen content (17.65 %), together with its boiling point, makes NH$_3$ among the best chemicals for its use as fuel or hydrogen carrier. This perspective makes of interest to develop novel efficient NH$_3$ preparation procedures to complement Haber-Bosch synthesis and that could be implemented on site without the need of a very large infrastructure.

This scenario has motivated a current interest in the electrochemical nitrate reduction reaction (NO$_3$RR), since this process could use renewable electricity and will be aligned with the current electrification of the chemical processes. Compared to N$_2$ reduction reaction, NO$_3$RR presents several advantages including higher Faradic efficiency and the use of soluble nitrate present in the electrolyte instead of insoluble N$_2$ gas that complicates considerably the electrolyzer design and diminishes substantially the process efficiency. In addition, kinetics and thermodynamic considerations also favor nitrate as substrate for NH$_3$ preparation in comparison to N$_2$. [1, 2] In addition, adapted nitrate reduction to NH$_3$ could also have some interest in some waste water streams from certain industries having high nitrate concentration, avoiding the negative environmental impact caused by eutrophication of these waters.

Selective NO$_3$RR to NH$_3$ is a complex process that requires the transfer of eight electrons and the corresponding number of H$^+$ through a multistep reaction mechanism that can result in the formation of several undesirable by-products. [3, 4] In addition, the efficiency of NO$_3$RR can be considerably decreased by the occurrence of concurrent hydrogen evolution reaction (HER). [5-7] Therefore, although there are several recent reports in the literature describing efficient NO$_3$RR, [8-12] there is still room for improvement regarding Faradic efficiency, easy to prepare electrocatalysts based on abundant elements that can operate at high current densities forming selectively NH$_3$ at high rates.

It is known that transition metal carbides may have d orbital center energy similar to those of noble metals, and that their unoccupied d orbitals can overlap with those of adsorbates promoting catalytic reactions on their surfaces. In this regard, it has been shown that Mo$_2$C is an efficient electrocatalyst for the N$_2$ reduction reaction to NH$_3$, due to the high N$_2$ adsorption capacity on this material combined with a catalytic hydrogenation activity. [13-15] In this regard, a common strategy to further increase the activity of electrocatalysts is doping with appropriate heteroatoms. [16, 17] Adequate doping can tune the electronic states on the catalyst surface, therefore, adjusting the adsorption strength of the catalyst to reactants and intermediates to the optimal values. As an example of doping in NO$_3$RR, it has been reported that Fe doping of Co$_3$O$_4$ increases the efficiency of NH$_3$ by enhancing NO$_3$-adsorption and inhibiting HER. [18] On the other hand, anchoring Fe atoms on Co$_3$O$_4$ avoids corrosion and surface oxidation of iron-based catalysts that are typically responsible for the poor stability of these electrocatalysts. [19, 20]

Inspired by this precedent, herein we report the electrocatalytic activity for NO$_3$RR of a Fe-doped Mo$_2$C catalyst. X-ray absorption measurements and spherical aberration-corrected transmission electron microscopy (AC-HAADF-STEM) with single atom resolution show that this Fe-doped Mo$_2$C material (FMC) contains single-atom Fe embedded into the Mo$_2$C lattice. The FMC catalyst exhibits excellent NO$_3$RR performance at low potential -0.3 V vs RHE, with NH$_3$ yield and FE reaching 14.66 mg$_{NH_3}$ h$^{-1}$ cm$^{-2}$ and 94.35 %, respectively. These excellent performance has been rationalized by DFT calculations that show that single atom Fe doping in the Mo$_2$C changes the reaction path, decreasing the potential barrier from 1.36 eV to 0.89 eV. In addition, Fe-doped samples facilitate the adsorption of intermediates and promote desorption of NH$_3$, thereby promoting the reduction of NO$_3^-$ to NH$_3$. Based on this FMC as electrode, a Zn-NO$_3^-$ battery, with a power density of 3.8 mW cm$^{-2}$ was developed. Therefore, our study shows a new synthesis method for single-atom doped electrocatalysts with considerable efficiency for NH$_3$ production.

## 2. Experimental Section

*2.1 Chemicals and reagents*

All chemical reagents were purchased commercially and employed without any further purification. Deionized water (MilliQ) was obtained via IQ 7000 purification system from Merck. Solvents such as ethanol (absolute, analytical grade), acetic acid (glacial, ReagentPlus, ≥ 99%) and acetone (≥ 99.5%) were purchased from Scharlau. Iron (II) chloride (99.99 % trace metals basis), chitosan (low molecular weight), ammonium molybdate tetrahydrate (NH$_4$)$_6$Mo$_7$O$_{24}$·4H$_2$O (ACS reagent, ≥ 98%), potassium hydroxide KOH (ACS reagent, ≥ 98%), zinc acetate (reagent grade, 98 %), ammonium hydroxide NH$_4$OH (ACS reagent, 28.0-30.0 % NH$_3$ basis), potassium nitrate KNO$_3$, carbon black (ACS reagent, ≥ 99%) were purchased from Sigma-Aldrich. Hydrogen gas (99.99 %) and argon gas (99.99 %) were purchased from Abello Linde SA.

*2.2 Materials preparation*

Commercially available reagents were purchased from Aldrich and used without further purification. Ultrapure MilliQ water was used in all experiments. Aqueous chitosan solutions were prepared by magnetic stirring for 30 min 0.3 g of this polymer in 50 mL ultrapure water adding 300 uL of CH$_3$COOH to assist dissolution. This solution was impregnated with (NH$_4$)$_6$Mo$_7$O$_{24}$*4H$_2$O (309 mg, 0.2 mmol) and FeCl$_2$ (177.5 mg, 1.4 mmol) by previously dissolving separately these two salts in 15 ml MilliQ water each with the assistance of ultrasounds until complete dissolution. Then the two metal salt solutions were dropped consecutively into the chitosan hydrogel solution. After 60 min stirring, the Mo-Fe impregnated chitosan hydrogel was place in an autoclave and submitted to hydrothermal reaction at 175°C for 20 h. The resulting suspension was filtered and the brown solid was thoroughly washed with MilliQ water and dried at 50 °C overnight. The resulting dry powder was put in a ceramic crucible in an electric tubular furnace under 100 mL/min Ar flow and pyrolyzed at 800 °C for 2 h, heating from the ambient temperature at a rate of 3 °C/min. The sample resulting from the pyrolysis was named as FMC. From this sample, FMC/CB was prepared by dispersing in 30 mL of MilliQ water with the aid of an ultrasound tip (770 W, 4 h) 50 mg FMC and 50 mg carbon black (CB). During the treatment the suspension was immersed in an ice bath to maintain the temperature. FMC/CB was obtained by filtration and dried in an oven at 50 °C overnight. Reference MC/CB and FC/CB samples were prepared following the same experimental procedure as for FMC/CB, except that no iron (MC/CB) or molybdenum (FC/CB) salts were used in the chitosan impregnation step.

*2.2 Characterization*

Powder X-ray diffraction (PXRD) patterns were recorded on a Shimadzu XRD-7000 diffractometer using Cu Kα radiation (λ=1.5418 Å), operating at 40 kV and 40 mA at a scan rate of 10° per min in the 2θ angle range between 2 to 90°. Transmission electron microscopy (TEM) images were acquired using a Philips CM300 FEG microscope operating at 200 kV, coupled with an X-Max 80 energy dispersive X-ray (EDX) detector (Oxford instruments). The electron microscope is equipped with the STEM unit and high-angle annular dark field (HAADF) image detectors. Ni and Mo analyses were measured by inductively coupled plasma-optical emission spectrophotometry (ICP-OES, Varian 715-ES, CA, USA). Specimens were prepared by depositing onto a carbon-coated copper TEM grid one micreodrop of the material previously dispersed in ethanol by sonication and allowing ethanol to evaporate at room temperature. $^1$H NMR spectra were recorded on a Bruker AV400 (400 MHz) spectrometer using known concentration DMSO as the internal standard. XPS data were acquired on an SPECS spectrometer equipped with a Phoibos 150 MCD-9 detector using a non-monochromatic X-ray source (Al) operating at 200 W. Before spectrum acquisition, the samples were evacuated in the prechamber of the equipment operating at $1*10^{-9}$ mbar pressure. Quantification of the atomic ratios of the elements was performed from the area of the corresponding peaks after nonlinear Shirley-type background subtraction and correcting by raw data with the relative response of each element in the spectrometer. The Fe K-edge X-ray absorption fine structure (XAFS) spectra were collected at BL11B beamline of Shanghai Synchrotron Radiation Facility (SSRF), China. The storage ring of SSRF was working at the energy of 3.5 GeV with an electron current of 240 mA in the top-up mode. The hard X-ray was monochromatized with Si (111) double-crystal monochromator and the detuning was done by 30 % to remove harmonics. The acquired EXAFS data were processed according to the standard procedures using the ATHENA module implemented in the IFEFFIT software packages. [21] The k2-weighted χ(k) data in the k-space ranging from 2.5-11.0 Å$^{-1}$ were Fourier transformed to real (R) space using a Hanning windows (dk = 1.0 Å$^{-1}$) to separate the EXAFS contributions from different coordination shells.

To obtain the detailed structural parameters around Fe atom in the as-prepared samples, quantitative curve-fittings were carried out for the Fourier transformed $k^2\chi(k)$ in the R-space using the ARTEMIS module of IFEFFIT. [22]

*2.3 Electrochemical measurements*

*2.3.1 Electrochemical measurements of nitrate reduction reaction ($NO_3RR$)*

Nafion membrane was activated in 5 wt% aqueous $H_2O_2$ solution at 80 °C for 1 h, and then in ultrapure water at 80 °C for another 1 h. All electrochemical measurements were performed under the Ar atmosphere. The electrocatalyst was previously deposited on a freshly polished glassy carbon by dispersing the solid material in ethanol containing 1 wt.% Paraloid as binder. The catalysts loading on the electrodes was adjusted to 1 mg/cm$^2$, approximately. The supported catalysts, Hg/HgO (1 M NaOH), Pt wire and 0.1 M KOH + $KNO_3$ were used as work electrode, reference electrode, counter electrode and electrolyte, respectively. LSV tests were conducted at a scan rate of 5 mV s$^{-1}$ in the three-electrode system. The chronoamperometry tests were conducted at various applied potentials in a H-type cell of 28 mL capacity with electrolyte separated by a Nafion 117 membrane. The voltage and current was set with a Gamry potentiostat (Interface 5000E). All potentials in this work refer to RHE, E(RHE) = E(Hg/HgO) + 0.14 + 0.059pH, where pH of the measurements was 13. The apparent activation energy (Ea) of $NO_3RR$ was calculated using the Arrhenius formula as follows: $i_k = Ae(-E_a/RT)$ where Ea is the apparent activation energy, $i_k$ is the kinetic current at different temperatures, A is the pre-exponential factor, T is the absolute reaction temperature and R is the universal gas constant.

*2.3.2 Aqueous rechargeable Zn-nitrate electrochemical cell*

Measurements regarding aqueous rechargeable Zn-nitrate battery were carried out in H-cell divided by an anion exchange membrane. The anode compartment was filled with an aqueous solution of $Zn(CH_3COO)_2$ (0.02 M) and KOH (1 M), while the cathode compartment was filled with an aqueous solution of 1 M $KNO_3$ with 1 M KOH. Catholyte was circulating in a closed system during experiments. A freshly polished zinc plate (1 * 2 cm$^2$) with sand paper was used as the anode. The discharging polarization curves were measured at a scan rate of 5 mV/s and galvanostatic tests were conducted using Gamry potentiostat (Interface 5000E) at room temperature.

The power density (P) of zinc-nitrate battery was determined by P = I×U, where I and U are the discharge current density and voltage, respectively. The electrochemical reactions in Zn-nitrate battery are considered to be the following:

Cathode reaction: $NO_3^- + 7H_2O + 8e^- \rightarrow NH_4OH + 9OH^-$

Anode reaction: $4Zn + 8OH^- \rightarrow 4ZnO + 4H_2O + 8e^-$

Overall reaction: $4Zn + NO_3^- + 3H_2O \rightarrow 4ZnO + NH_4OH + OH^-$

*2.4. Product quantification*

The evolving gases were analyzed with an Agilent 490 Micro GC system (Molsieve 5 Å column using Ar as carrier gas). N-containing species were detected by colorimetry with ultraviolet-visible (UV-Vis) spectrophotometer. In particular, $NH_3$ was also quantified by the solution $^1H$ NMR spectroscopy method. The specific detection methods were presented in Supporting information.

*2.5 Computational method*

All calculations were performed using spin-polarized density functional theory (DFT) as implemented in the Vienna ab initio simulation package (VASP). [23-25] The Perdew-Burke-Emzerhof (PBE) functional with a plane-wave cutoff energy of 500 eV was used. For structural optimizations, the Brillouin zone was sampled using a 5 × 5 × 1 k-point grid based on the Monkhorst-Pack scheme, centered at Gamma. A vacuum space of 1.5 nm in the *z* direction (perpendicular to the basal plane) was used to avoid interactions between periodic images. The convergence criteria for the force on each atom was set to 0.02 eV/Å, while the electronic structure energy convergence criteria was $10^{-5}$ eV. The Grimme D3 method with Becke-Johnson parameters [26] were employed to account for Van der Waals interactions. [27] The vibrational modes were calculated at 298.15 K to obtain the zero-point energy, entropy, and temperature corrections to enthalpy (their contributions is listed in Table S4-S6). More details are presented in Supporting information.

## 3. Results and Discussion

### 3.1. Materials preparation and characterization

The synthetic strategy for preparing the Fe-doped $Mo_2C$ (FMC) catalyst is described in detail in the Supporting Information. First, an aqueous solution of equimolar amounts of $(NH_4)_6Mo_7O_{24}$ and $FeCl_2$ metal salts n 15 ml of MilliQ water was prepared and added into another aqueous solution of chitosan dissolved with the aid of some acetic acid. The mixture was heated in an autoclave at 175 ºC for 20 h to obtain a hydrogel. Subsequently, the resulting hydrogel containing well-dispersed metal ions into the chitosan fibrils were submitted to pyrolysis at 900 ºC, for 2 h. Finally, FMC/CB were obtained by ultrasonicating the mixed solution of FMC and carbon black (CB) for 4 h. The role of CB is to increase the electrical conductivity of the electrocatalyst layer and to act as binder. Inconsistent electrochemical measurements were observed in the absence of CB additive due to electrocatalyst detachment. Figure 1a illustrates the preparation procedure.

The XRD of the relevant sample is shown in Figure 1b. FMC is composed of a mixture of three separate phases corresponding to $Fe_3Mo_3C$, $Mo_2C$ as major components and Fe metal as very minor component. Unexpectedly, the sonication process to introduce carbon black (CB) allows a spontaneous crystal phase separation, the resulting FMC/CB containing only $Mo_2C$ according to PXRD. No Fe-containing species could be detected in the PXRD of the resulting FMC/CB. Furthermore, careful inspection of the XRD (Fig. 1c) shows that the main peaks of $Mo_2C$ in the 34-41° undergo a shift to higher angles of +0.05° in FMC and this shift further increases to +0.1° in the resulting FMC/CB. These shifts in PXRD are indicating the occurrence of a gradual doping that incorporates an heteroatom in the structure of $Mo_2C$. [28] Therefore, a reasonable hypothesis to explain these PXRD data is that Fe atoms from the disappearing $Fe_3Mo_3C$ phase are partially incorporated into the $Mo_2C$ lattice due to the action of the extended periods of high-power ultrasounds (700 W, 4 h) that will cause also the dissolution of $Fe_3Mo_3C$ in the deionized water even at low temperature (0 ºC). Chemical analysis by ICP-OES of the aqueous phase after the sonication and the resulting FMC/CB powder is consistent with this proposal with most of the Fe being present in solution, but also in FMC/CB (Table S1). The PXRD patterns of FC/CB, FC, MC/CB and MC have been also measured, and they are presented as Fig. S1.

The morphology of the samples was analyzed by TEM and SEM. TEM image of FMC show that the solid is composed of agglomeration of particles (Fig. S2). Note that N doping derives from chitosan precursor composition and should not play any major role in the process compared to the electrocatalyst. SEM images of FMC and FMC/CB show insignificant changes in morphology (Fig. S3). The TEM image (Fig. 1d) of FMC/CB shows a homogeneous distribution of nanoparticles of 4.8+1.2 nm average size on the carbon support. The HRTEM image (Fig. 1e) shows a crystalline lattice spacing of 0.225 nm, which can be assigned to the (101) plane of the Mo$_2$C crystal, although with somewhat shorter distance than the reference one of 0.230 nm. This change in interplanar spacing can be attributed to the shrinking of Mo$_2$C lattice by the presence of Fe atoms and it is fully consistent with the shifts in the XRD diffraction peaks measured for FMC/CB. The TEM images and nanoparticle size distribution of FC/CB and MC/CB are provided in Fig. S4-5.

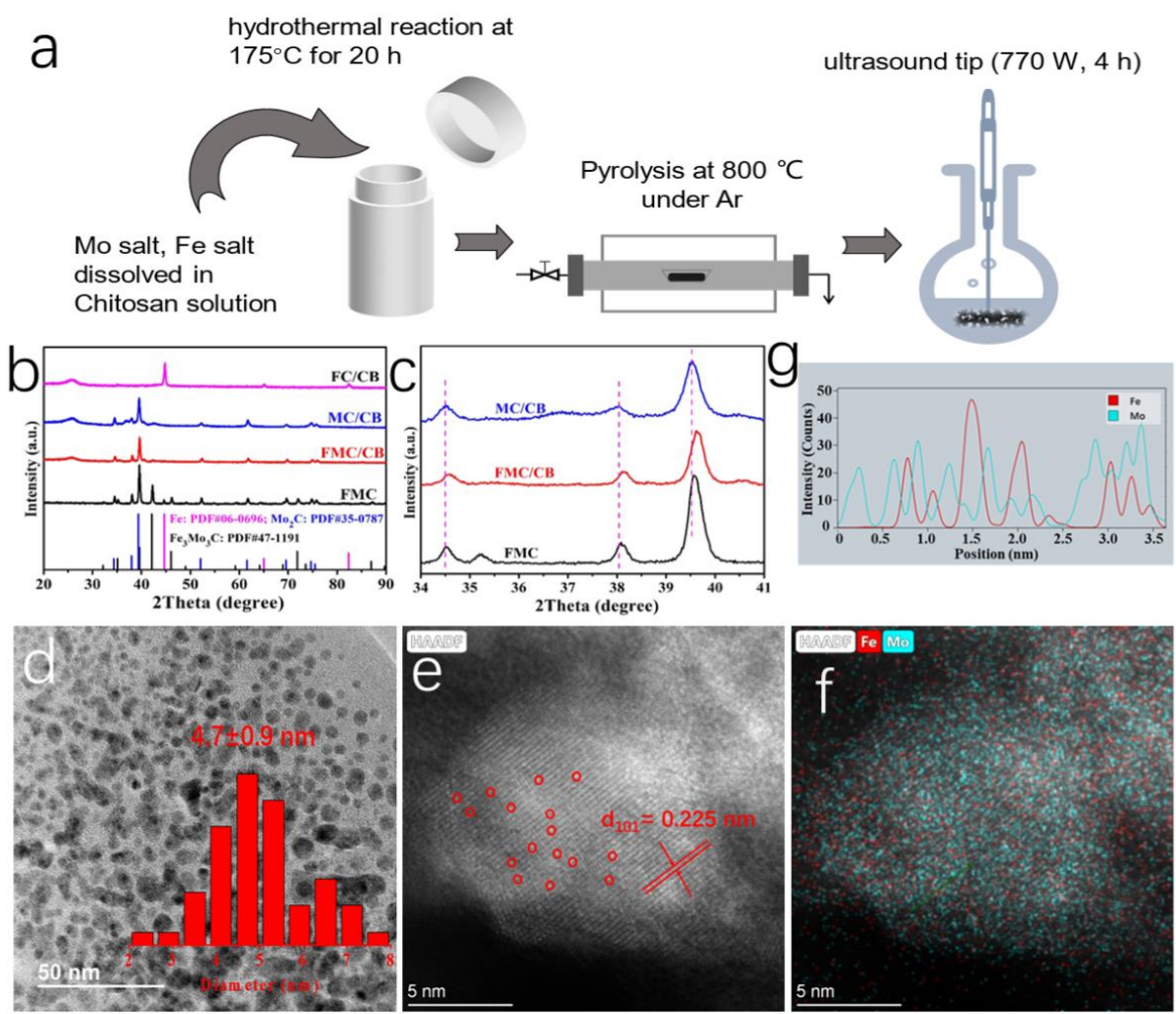

**Figure 1.** (a) Schematic diagram showing the preparation of FMC/CB catalyst; XRD patterns (b) and zoom in of 30-40 ° zone (c) of samples under study; (d) TEM image of FMC/CB and corresponding histogram of the particle size distribution; (e) HAADF-STEM images of FMC/CB indicating the interplanar distance and circling Fe atoms and (f) STEM-EDS mapping of Fe (red) and Mo (green) in FMC/CB; (g) The line profiles of Fe and Mo obtained from the HAADF-STEM image.
9

The Fe dispersion in the FMC/CB catalyst at the level of single atom can be confirmed by the high-angle annular dark-field scanning transmission electron microscopy (HAADF-STEM) image with atomic resolution recorded for the $Mo_2C$ particles (Fig. 1e). Its corresponding line profiles for HAADF intensity analysis (Fig. 1g) present that the Fe atoms are dispersed by replacing the Mo atomic positions or being located between the $Mo_2C$ (101) planes. Notably, the images show the absence of Fe nanoparticles or clusters in the FMC/CB sample and the atomic dispersion of Fe element can be observed in the scanning transmission electron microscopy energy-dispersive X-ray spectroscopy (STEM-EDS) (Fig. 1f and Fig. S6).

The valence state of Mo and Fe can be determined by X-ray photoelectron spectroscopy (XPS) analysis. Mo3d of MC/CB can be deconvoluted in two sets of peaks with binding energies of 228.7 and 232.7 eV (Fig. 2a), which are attributed to Mo carbide and oxidized Mo species, respectively. It is worth noting that the binding energy of the Mo carbide peak of FMC/CB is shifted towards lower values from 228.7 to 228.25 eV, indicating that Fe atoms are giving electron density to the Mo atoms. Accordingly, the Fe2p peak of FMC/CB exhibits a slight shift towards higher binding energy values compared to FC/CB, due to the electron density donation of Fe atoms to Mo (Fig. S7). This effect of Fe on Mo in XPS has been previously reported. [29, 30]

The Fe K-edge X-ray absorption near-edge structure (XANES) measurements of FMC and FMC/CB were further performed to investigate the oxidation number and coordination states. FMC features a series of characteristic absorption peaks similar to those of $Fe_7C_3$ [31] at 7112 and 7130-7150 eV (Fig. 2b), indicating that the Fe local structure in FMC closely resembles that of Fe carbides. Furthermore, the Fourier-transformed (FT) spectrum of FMC (Fig. S8-9) reveals prominent coordination peak at 1.44 Å associated with the first-shell Fe-C scattering, which is consistent with the XANES results. Regarding the second FT peak at 2.22 Å, we resorted to quantitative curve-fitting analysis to infer Fe-Fe coordination number N= 2.4 at an interatomic distance of R= 2.52 Å and Fe-Mo coordination number N= 0.7 at an interatomic distance of R= 2.84. This suggests that most of the iron is present in islands of Fe carbides, whereas truly mixed FeMo carbides are less populated.

In comparison with FMC, the $k^2\chi(k)$ oscillation curve of FMC/CB attenuates noticeably slower as it approaches higher k value in the k-space (Fig. 2c-d). Rigorous fitting analysis determines a well-defined Fe-Mo contribution in the nearest neighbor of the Fe center. Furthermore, the indicated Fe-Mo bond length of 2.96 Å is equivalent to the Mo-Mo bond length (2.95-3.01 Å) of $Mo_2C$ (PDF#35-0787). Table S2 collects all the XANES fitting parameters for FMC and FMC/CB. Taking all these data into account, it is inferred that Fe progressively substitutes Mo sites in the Mo$_2$C structure as the number of sonication steps in sample preparation increases.

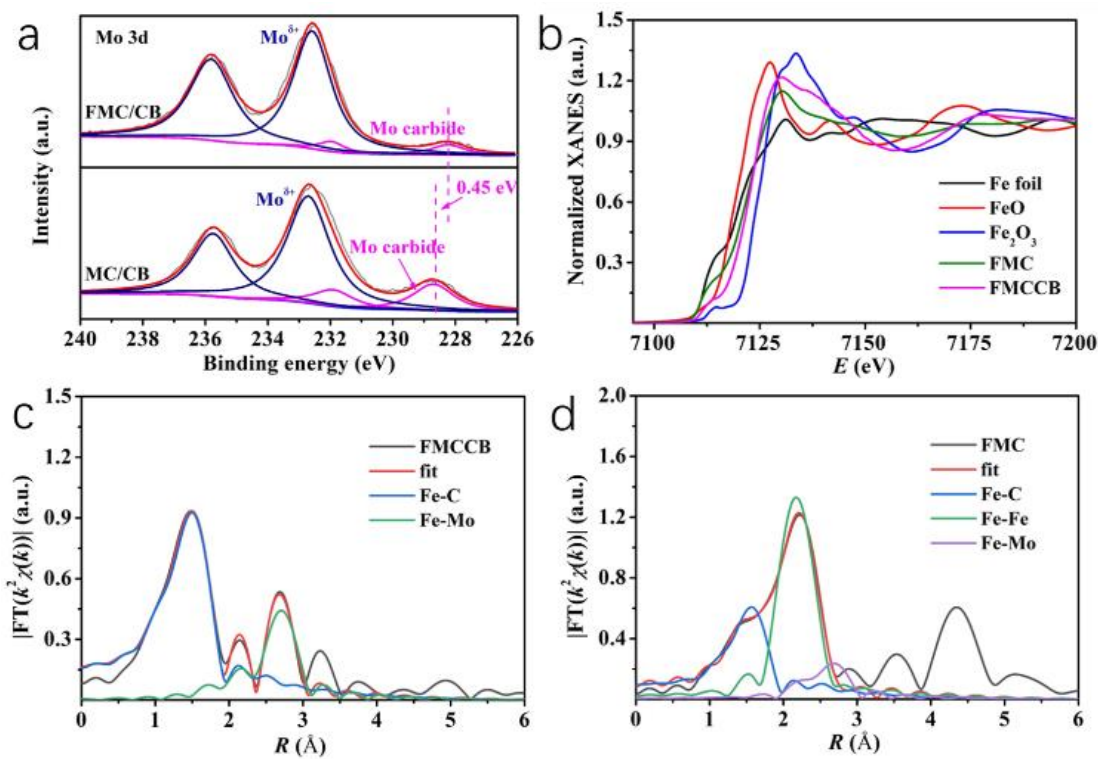

**Figure 2**. (a) XPS Mo3d spectra and the best fitting to individual component. The shift in the peak maximum of Mo carbide due to the influence of Fe on the lattice has been highlighted; (b) Fe K-edge XANES spectrum of FMC, FMC/CB, FeO, Fe$_2$O$_3$, Fe foil; The EXAFS fitting curves FMC/CB at R space (c) and k space (d).

## 3.2. Electrocatalytic NO$_3$RR performance

In light of these evidences, we carried out a series of electrochemical measurements to determine the electrocatalytic NO$_3$RR performance of the samples in a nitrate-containing electrolyte. The electrochemically active surface area (ECSA) was initially estimated based on measurements of the double-layer capacitance. As shown in Fig. S10, FMC/CB exhibits a larger ECSA C$_{dl}$ of 28.4 mF/cm$^2$ than the other samples. Furthermore, linear sweep voltammogram (LSV) plots shown in Fig. 3a indicate that all samples show a significant current density increase when the electrolyte contains NO$_3^-$ in comparison with inert electrolyte. Especially, FMC/CB incorporating Fe atoms into the Mo$_2$C lattice displayed the highest current density at low potential (more than 50 mA/cm$^2$ at -0.3 V vs RHE).

We further conducted the NO$_3$RR experiment in an H-type cell saturated with Ar gas to achieve an enhanced efficiency and NH$_3$ yield. Based on the previous LSV data, chronoamperometry (CA) measurements were carried out at various applied potentials to quantitatively assess the NO$_3$RR performance of the samples determining NH$_3$ yields and FEs. The results are presented in Figure 3b that compares the performance of FMC/CB, FC/CB and MC/CB for the NO$_3$RR in the potential range from −0.1 to −0.5 V. Obviously, FC/CB exhibits a relatively high FE of by-product NO$_2^-$ at low potential, while the occurrence of HER can be effectively avoided at high potential. In contrast, although MC/CB exhibits lower FE of NO$_2^-$ at low potential, a large proportion of by-product H$_2$ is produced. Interestingly, FMC/CB just combines the advantages of FC/CB and MC/CB without their disadvantages, exhibiting the highest NH$_3$ FE over a wide potential range. In addition, the onset potential of FMC/CB was as low as -0.1 V vs RHE, and showed the highest FE of 94.35% at -0.3 V, with the corresponding NH$_3$ yield reaching 14.66 mg h cm$^{-2}$ (Fig. 3c). It is worth mentioning that the FE of NH$_3$ formation was always higher than 85% in the potential window of the measurement. In addition to using colorimetric methods for quantitative analysis of NH$_3$ (Fig. S11 and S12), $^1$H nuclear magnetic resonance (NMR) spectroscopy was also used to quantify NH$_3$ (Fig. S13). The coincidence between the two quantification methods was remarkable, thus re-confirming the validity of the experimental data. The comparison of the electrocatalytic NO$_3$RR performance here measured for FMC/CB with those of reported electrocatalysts is summarized in Table S3. This literature survey clearly reveals that FMC/CB exhibits superior NH$_3$ yield and FE than those electrocatalysts previously reported in the literature.

Considering that industrial wastewaters may differ in a wide range of NO$_3^-$ concentrations, FMC/CB was also screened as electrocatalyst using electrolytes containing different NO$_3^-$ concentrations (Fig. S14). The results show that FMC/CB exhibits good electrocatalyst performance for NO$_3$RR to NH$_3$ in range 5-100 mM of KNO$_3$ solutions. The apparent activation energy (Ea) of NO3RR was calculated (Fig. 3d) from the Arrhenius plot using the formula: i$_k$ = Ae(-Ea/RT). As shown in Figure 3d, FMC/CB shows the lowest apparent activation energy of 10.1 kJ mol$^{-1}$.

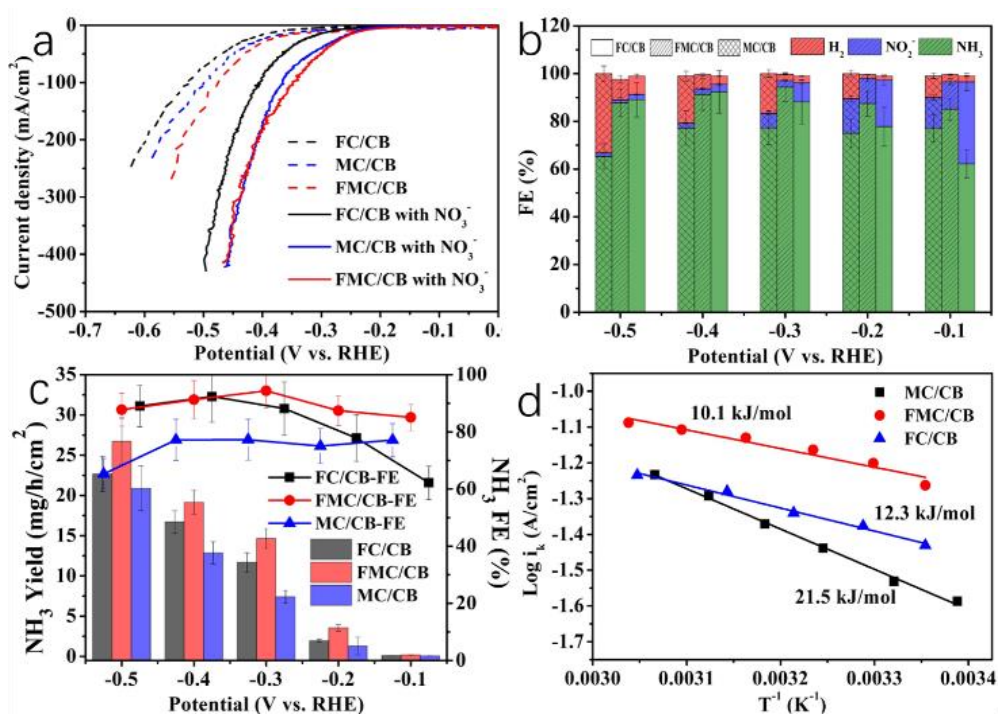

**Figure 3**. (a) LSV of different samples with or without $NO_3^-$; (b) $NO_3RR$ performance of different samples at -0.1- -0.5 V vs RHE; (c) $NH_3$ yield and corresponding FE of the samples under study; (d) apparent activation energy (Ea) of different samples.

Stability is also a relevant parameter for application. For that reason, we have performed a stability test to FMC/CB, monitoring the current density and $NH_3$ FE upon continuous -0.3 V vs. RHE applied bias for 24 h. As can be seen in Fig. 15 in SI, nearly constant current density and $NH_3$ FE were observed. Moreover, the LSV of the FMC/CB electrode prior and after the stability test is also presented as Fig. S15 b, demonstrating a notable stability during this operation for this time period. In order to further highlight the stability of FMC/CB, the XRD patterns of FMC/CB after the stability test has been also recorded and compared with that of the fresh sample (Fig. S16 in SI). As can be seen, the crystal structure of FMC/CB remains unchanged after -0.3 V vs. RHE applied bias for 24 h.

### 3.3. FMC/CB as electrode component of a rechargeable Zn-$NO_3^-$ battery

After confirming the excellent performance of FMC/CB as electrocatalyst for $NO_3RR$ to ammonia, this material was used as electrode in an exploratory study of its suitability for a novel rechargeable Zn-$NO_3^-$ battery by anchoring FMC/CB on a carbon paper as the cathode and assembling it to zinc foil as the anode. The aim was to investigate the efficiency in which electrons coming from the NO$_3$RR process can be utilized in a nitrate-based battery, while still producing value-added NH$_3$ production. Fig. 4a shows the charge-discharge LSV curves of FMC/CB-based Zn-nitrate battery. The discharging curve for such Zn-NO$_3$-battery shows an increased output current density with the cathodic potential being more negative. Fig. 4b-c shows the discharge curve of the Zn-NO$_3^-$ battery, as well as the NH$_3$ yield and corresponding FE to NH$_3$ under different discharge current density from 1 to 20 mA/cm$^2$. The power density of Zn-NO$_3^-$ battery reaches the peak of 3.8 mW cm$^{-2}$. The FMC/CB based Zn-NO$_3^-$ battery delivers a high NH$_3$ formation rate of 1.8 mg/h/cm$^2$ at 20 mA/cm$^2$ and achieves a high FE of 87.7%. Fig. 4d shows discharge–charge cycles of Zn-nitrate battery at a constant current density of 10 mA/cm$^2$, showing a notable stability. In addition, it should be noted that the Zn-nitrate battery can be rechargeable. The anode involves the dissolution and deposition of Zn and the cathode includes NO$_3$RR and OER during discharge and charge process, respectively. Therefore, Zn-NO$_3^-$ battery based on FMC/CB as electrode has potential applicability, further expanding the field of Zn-based battery.

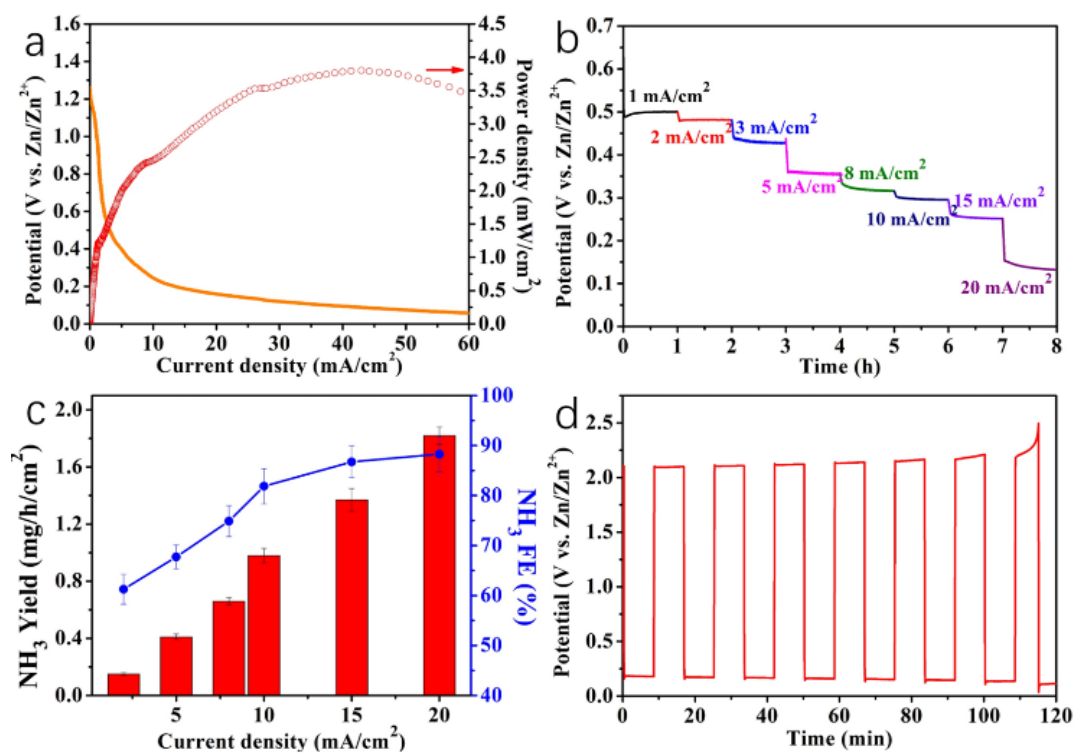

**Figure 4**. (a) Discharging current-voltage curve of the Zn-NO$_3^-$ battery; (b) Discharging tests at various current densities. The same electrode was used in these measurements; (c) NH$_3$ yield (red bars, left axis) and FE values towards NH$_3$ formation (blue line, right axis) at different current densities of the Zn-NO$_3^-$ battery; (d) The discharge–charge processes of Zn-NO$_3^-$ battery at a constant current density of 10 mA/cm$^2$.

## 3.4. DFT calculations

DFT simulations were performed to elucidate the NO$_3$RR mechanism, and to investigate the origin of the difference in the activities of MC, FC and FMC. On pure MC, the reaction starts with the adsorption of *NO$_3^-$, which is favorable, the anion resulting strongly bound, with a calculated adsorption energy of -1.14 eV (Fig. 5a). Then, a first protonation occurs forming *NO$_3$H which is more stable than the precursors by 0.16 eV. Water elimination leads to the formation of *NO$_2$, which is strongly stabilized, lying at -2.89 eV. Protonation of this intermediate leads to *NO$_2$H formation, which is the potential-dependent step (PDS) of the whole reaction, of which the maximum reaction free energy is 1.36 eV at -0.33 V vs standard hydrogen electrode (SHE), followed by the subsequent exergonic water elimination to obtain *NO, which is strongly stabilized, with an energy of -3.48 eV. From the *NO intermediate, two pathways can be followed, namely either the protonation of the nitrogen atom leading to *NHO or the protonation of the oxygen atom leading to *NOH. Both pathways are thermodynamically uphill, requiring an energy of 1.15 and 1.25 eV to be formed, respectively. From *NOH, the subsequent water elimination to obtain *N is strongly exergonic, at -3.74 eV. The subsequent protonation to *NH and *NH$_2$ are isoenergetic, and the final protonation to *NH$_3$ is exergonic. On the other hand, the formation of *NHO leads to the protonation of the oxygen atom to obtain *NHOH, from which water elimination can occur to obtain *NH and eventually *NH$_3$, which is strongly adsorbed to the surface with an energy of -1.23 eV.

Introducing Fe in the MC structure (FMC) results to three major changes in the pathway for ammonia reduction (Fig 5a). First, the *NO$_2$H intermediate (which was the PDS for the undoped surface) has not been found to be stable on this interface. Secondly, the presence of Fe in the structure moves the PDS from the previous *NO$_2$H formation to the generation of the *NHO intermediate from *NO, the last species having an energy value of 0.89 eV. Finally, while on the pristine surface both pathways from either *NOH or *NHO could be followed due to their close energies, now the *NHO pathway is the most probable, since this intermediate has been found to have a lower energy compared to the alternative *NOH possibility with an energy difference of 0.5 eV, making the *NOH pathway no longer available. Thus, from *NHO a second protonation and subsequent water elimination affords *NH, from which strongly exergonic steps lead to the formation of *NH$_3$.

The presence of Fe as dopant has also a strong effect on the adsorption energy of NO$_3^-$ and NH$_3$, with values that are almost halved compared to the pristine surface, namely -0.72 eV for *NO$_3$ and -0.55 eV for *NH$_3$. The much lower adsorption energy obtained for NH$_3$ might hint to its ease of desorption on this surface compared to the pristine case.

On the other hand, when we consider the FC surface we observe an intermediate situation compared to MC and FMC, in agreement with the experiments. Three major differences can be seen for this surface. Specifically, the *NO$_3$H energy is now the less favorable among the three surfaces, with value of 0.12 eV (compared to 0.05 eV and -0.16 eV for FMC and MC, respectively), suggesting a more sluggish reaction. This can be rationalized considering the high NO$_3^-$ adsorption energy, of -2.35 eV. Secondly, the *NO$_2$H intermediate is once more observed, as for the MC surface, but with an energy value difference from *NO of 0.73 eV which, albeit lower than for MC, still requires a high energy input to be overcome. Finally, the formation of the *NHO intermediate from *NO is now the PDS of the whole pathway, with an energy difference value of 0.95 eV, higher than for FMC by 0.06 eV. As for FMC, the *NOH pathway from *NO oxygen protonation is not accessible, due to the high energy required to obtain the *NOH intermediate. Interestingly, the protonation of *NHO to *NHOH leads to the isoenergetic dissociation of the species into *NH and *OH on the surface, making this step less favorable than on both FMC and MC surfaces. The comparison of reaction pathways, for MC (a) FMC (b) and FC (c) interfaces at different applied potentials of 0 and -0.33 V, were shown in Fig. S17.

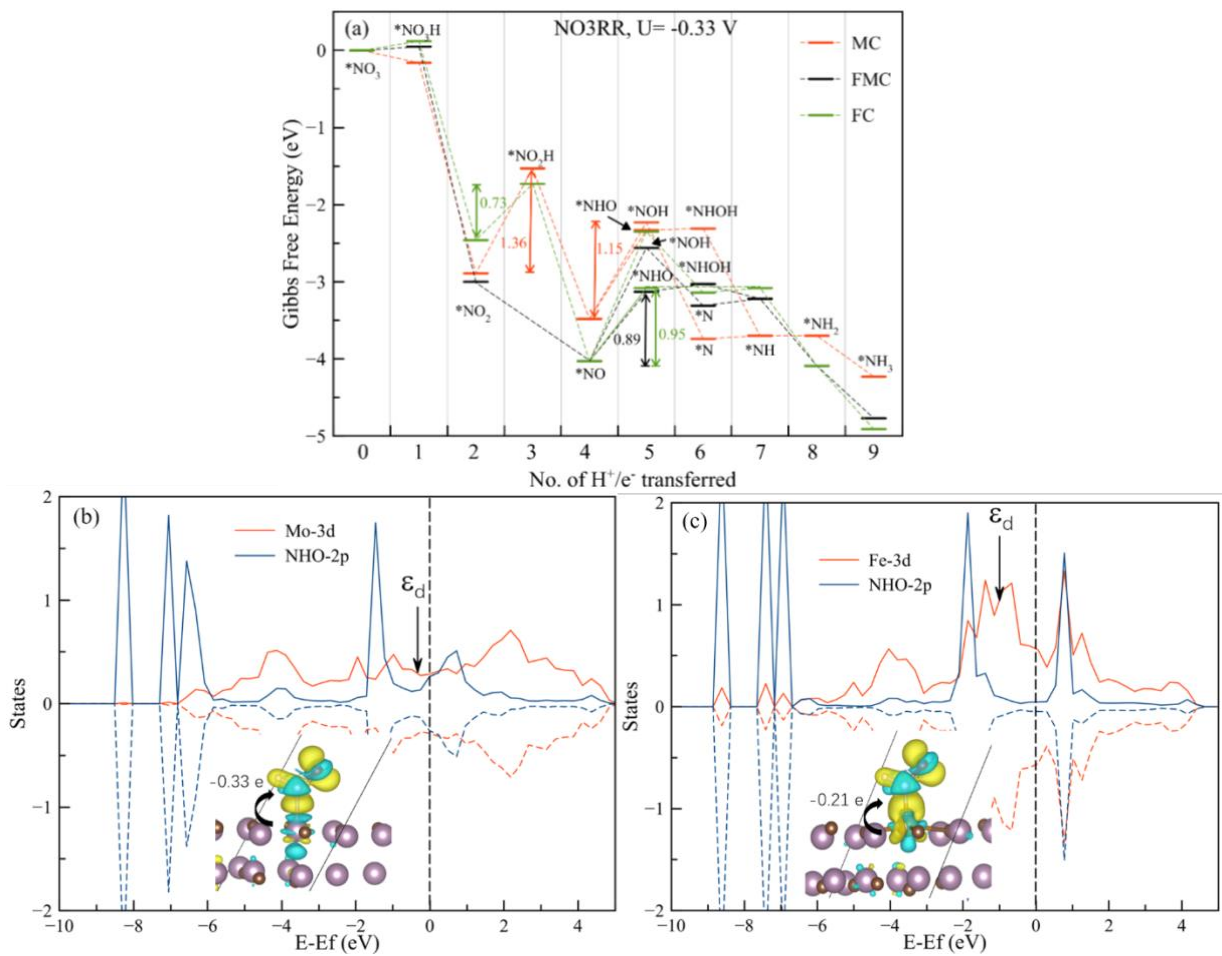

**Figure 5**. (a) Gibbs free energy reaction pathway for the NH$_3$ formation on MC, FMC and FC interfaces, at an applied potential of -0.33 V; Density of states (DOS) on the 3d orbitals for Mo (in MC) (b) and Fe (in FMC) (c) metal centres and on the 2p orbitals of NHO.

The experimental enhancement in electrocatalytic activity observed for the FMC surface can, thus, be explained as follows. The PDS of the whole pathway changes when Fe is present, leading to a decrease in barrier to be overcome from 1.36 eV to 0.89 eV going from MC to FMC. Moreover, if we compare now the two *NHO intermediates, it is observed that while when Fe is present the energy required to reach this intermediate is 0.89 eV, on the MC surface the energy is increased to 1.15 eV and on FC is 0.95 eV. This is a direct result of enhanced adsorption caused by the upshifted d-band center for the pristine MC surface (Fig. 5b-c).

In fact, the d-band energy center has been found at -0.98 eV for the FMC system, and is shifted up to -0.31 eV for the pristine MC surface. The main idea is that the closer to the Fermi energy the metal d-band center is, the stronger the adsorbate-metal interaction is, due to a lower occupation of antibonding states. This is also confirmed by the Bader charge analysis, from which the charge transfer (CT) for selected adsorbed intermediates can be obtained. For the

*NHO adsorbate, CT is larger for the pristine MC surface (-0.33 e) compared to the FMC system (-0.21 e), which rationalizes the higher activity for $NH_3$ formation of the latter.

Finally, we considered the hydrogen reduction reaction (HER) over the three surfaces. For all three surfaces, the HER is unfavorable compared to the NO3RR. In fact, we observe Gibbs free energy of 1.03, 0.76 and 0.39 eV for the FC, FMC and FC respectively, suggesting that there is little or no competition between the two reactions, and the favored process is the ammonia production.

## 4. Conclusion

In this work, we successfully incorporate Fe in the structure of $Mo_2C$ forming a really randomly distributed Fe-Mo carbide (FMC). The resulting FMC exhibits an excellent electrocatalytic activity for the production of ammonia via direct electroreduction of nitrate. The process is successful over a wide concentration range. FMC reaches a $NH_3$ yield of 14.66 mg h $cm^{-2}$ with an FE of 94.35% at low potential -0.3 V vs RHE. DFT calculations show that the presence of Fe uniformly distributed in the $Mo_2C$ lattice changes the relative energy of the reaction intermediates and it is responsible for a diminution of the potential barrier to be overcome from 1.36 eV for pristine MC to 0.89 eV for FMC. In addition, the presence of Fe in the $Mo_2C$ structure facilitates the adsorption of intermediates and promote the desorption of $NH_3$, thereby accelerating the reduction of $NO_3^-$ to $NH_3$. This work provides a new synthesis method for the preparation of mixed Mo carbides and opens up new avenues for collaborative site catalysts. In addition, the synthesized FMC is a suitable cathode for a novel rechargeable $Zn-NO_3$ battery in which in the discharge process the electrons of NO3RR are used to generate electricity, while producing ammonia at the same time, reaching a power density of 3.8 mW/$cm^2$ with a FE to $NH_3$ of 88 %. In the recharge process FMC promotes the OER. The present work first uncovers a convenient sonochemical process to prepare truly mixed Fe-Mo carbide and it shows the excellent electrocatalytic activity of the mixed bimetallic Fe-Mo carbide for $NH_3$ production, even broadening the field of zinc-based batteries.

**Declaration of Competing Interests**

The authors declare that they have no known competing financial interests or personal relationships that could have appeared to influence the work reported in this paper.

**Supplementary data**

Supplementary data associated with this article can be found in the online version. Supporting Figs. and Tables mentioned in the text.

**Acknowledgements**

Financial support by the Spanish Ministry of Science and Innovation (CEX-2021-001230-S and PDI2021-0126071-OB-CO21 funded by MCIN/AEI/ 10.13039/501100011033) and Generalitat Valenciana (Prometeo 2021/038 and Advanced Materials programme Graphica MFA/2022/023 with funding from European Union NextGenerationEU PRTR-C17.I1). J.H. thanks the Chinese Scholarship Council for a postgraduate scholarship to perform his PhD at Valencia. S.O. thanks the National Science Centre, Poland (grant no. UMO/2020/39/I/ST4/01446) and the "Excellence Initiative – Research University" (IDUB) Program, Action I.3.3 – "Establishment of the Institute for Advanced Studies (IAS)" for funding (grant no. UW/IDUB/2020/25). The computation was carried out with the support of the Interdisciplinary Center for Mathematical and Computational Modeling at the University of Warsaw (ICM UW) under grants no. G83-28 and GB80-24.

A mixed (Fe,Mo)$_2$C is an efficient electrocatalyst to promote the selective reduction of nitrate to NH$_3$ with a yield of 14.66 mg$_{NH_3}$ h$^{-1}$ cm$^{-2}$ and an FE of 94.35 % at low potential -0.3 V vs RHE. This material can also be used as cathode for an aqueous Zn-nitrate battery

Jiajun Hu, Silvio Osella, Eduardo Arizono dos Reis, Anelisse Brunca da Silva, Caue Ribeiro, Lucia Helena Mascaro, Josep Albero and Hermenegildo Garcia*

**Mixed Fe-Mo carbide prepared by a sonochemical synthesis as highly efficient nitrate reduction electrocatalyst**

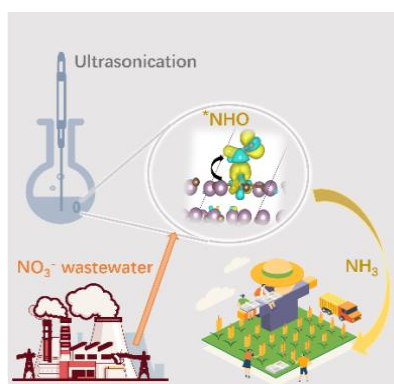